\documentclass[12pt]{article}
\usepackage{amssymb,amsmath,epsfig}

\numberwithin{equation}{section}

\begin{document}

\title{\textbf{Charged Vector Particles
Tunneling From 5D Black Hole and Black Ring}}

\author{Wajiha Javed$^1$ \thanks{wajiha.javed@ue.edu.pk;wajihajaved84@yahoo.com},
Riasat Ali$^1$ \thanks{riasatyasin@gmail.com} and G.
Abbas$^2$ \thanks{abbasg91@yahoo.com;ghulamabbas@iub.edu.pk}\\
$^1$Division of Science and Technology, University\\ of Education,
Township Campus, Lahore-54590, Pakistan\\
$^2$Department of Mathematics, The Islamia\\
University of Bahawalpur, Bahawalpur, Pakistan.}

\date{}
\maketitle

\begin{abstract}
In this paper, we investigate the Hawking radiation process as a
semiclassical quantum tunneling phenomenon from black ring and
Myers-Perry black holes in 5-dimensional (5D) spaces. Using
Lagrangian of Glashow-Weinberg-Salam model with background
electromagnetic field (for charged W-bosons) and the WKB
approximation, we will evaluate the tunneling rate/probability of
charged vector particles through horizons by taking into account the
electromagnetic vector potential. Moreover, we investigate the
corresponding Hawking temperature values by considering Boltzmann
factor for both cases and analyze the whole spectrum generally.
\end{abstract}

\section{Introduction}

In the early universe, by assuming the quantum back ground, Hawking
observed that black hole (BH) emits particles and the spectrum of
the passed off particles is purely thermal. Hawking expressed the
emission spectrum for all kinds of particles e.g. neutrinos,
photons, gravitons, electrons, positrons etc. and accounted the
emission rate for these particles. The vector particles or vector
bosons (with spin equals to $1$), i.e., $W^{\pm}$ and $Z$ play a
fundamental role as important component of the standard model for
electroweak interaction, so that the emission of such particles
should also be of more significance in the analysis of Hawking
radiation. Different methods have been suggested for the analysis of
Hawking radiation as a tunneling spectrum across the horizons of
BHs. The tunneling method is based on the fundamental physical
particles action which allows BH radiation.

Existence of the positive and negative energy pair of virtual
particles is similar to the existence of particle and anti-particle
pair creation, respectively. At instant, the negative and positive
energy virtual particles are created and annihilated in form of
pairs. The positive energy virtual particle disappears by tunnel
through the horizon and it is emitted as a depart of the Hawking
radiation. While, the negative energy particle goes inside the BH or
absorbed by the BH. The rule of conservation of energy is adapted in
the procedure. The specific value of temperature at which particles
emitted from the BH is known as Hawking temperature. For the
Schwarzschild back ground geometry the Hawking temperature is
obtained for scalar particle emission.

Emparan and Recall \cite{R1} discussed a rotating 5D black ring. Lim
and Teo \cite{R2} calculated the particles motion in the spacetime
of the rotating black ring. Chen and Teo \cite{R3} discussed
rotating black rings on Taub-NUT space by applying the
inverse-scattering procedure. Matsumoto, Yoshino and Kodama
\cite{R4} talked about the loss of BH's mass and angular momenta
through Hawking radiation. Cvetic and Guber \cite{R5} discussed
R-charged BHs (particular BHs in gauged super-gravity in $D=5$ and
$D=7$). Aliev \cite{R6} considered that a BH may posses a small
electric charged and constructed a 5-vector potential for
electromagnetic field in the background of Myers-Perry metric.
Sharif and Wajiha \cite{R7} have computed the Hawking radiation
spectrum for charged fermions tunneling from charged accelerating
and rotating BHs with NUT parameter. Jan and Gohar \cite{R8} applied
the Hamilton-Jacobi method to the Klein-Gordon equation by using WKB
approximation. They determined the tunneling probability of outgoing
charged scalar particles from the event horizon of BH and also
detect the corresponding Hawking temperature. Kruglov \cite{R9}
studied the Hawking radiation of spin-1 particles from BHs in
(1+1)-dimensions by using Proca equation. The procedure is viewed as
quantum tunneling of bosons through the event horizon. They
established the result that the emission temperature with the
Schwarzschild background geometry is similar to the Hawking
temperature of scalar particles emission. Kruglov also found the
Hawking temperature for the de Sitter, Rindler and Schwarzschild de
Sitter spacetimes. Li and Zu \cite{R10} have studied the tunneling
rate and Hawking temperature of scalar particles for
Gibbons-Maeda-Dilation BH.

Feng, Chen and Zu \cite{R11} analyzed the Hawking radiation of
vector particles from 4D and 5D BHs. Saleh, Thomas and Kofane
\cite{R12} obtained the Hawking radiation from a 5D Lovelock BH by
applying the Hamilton-Jacobi procedure. Saifullah and Yau \cite{R13}
studied the outgoing and ingoing probabilities and temperature of
the spin-$\frac{3}{2}$ particles. Chen and Huang \cite{R14} studied
the Hawking radiation of vector particles for Vaidya BHs. Li and
Chen \cite{R15} looked into vector particles tunneling (uncharged
and charged bosons) from the Kerr and Kerr-Newman BHs. Singh, Meitei
and Singh \cite{R16} have used the WKB approximation and
Hamiliton-Jacobi ansatz to the Proca equation and calculated the
tunneling rate and Hawking temperature of vector bosons for
Kerr-Newman BH. In this paper, the Hawking temperature is discussed
for three coordinate systems. Gursel and Sakalli \cite{R17} studied
the Hawking radiation of massive vector particles from rotating
warped anti-de Sitter BH in 3D and showed that the radial mapping
gives the tunneling rate for outgoing particles. Different authors
\cite{R18,R19} have found the Hawking temperature for various types
of particles from BHs. Li and Zhao \cite{R20} analyzed the massive
vector particles tunneling and found the Hawking temperature from
the neutral rotating anti-de Sitter BHs in conformal gravity by
applying tunneling procedure. Ali and Jusufi \cite{R21} studied the
quantum gravity effects on the Hawking radiation of charged spin-1
particles for noncommutative charged BHs, RN BHs and charged BHs.

It is to be observed that the kinetics of charged vector particles
are ruled along the Proca field equations by using WKB approximation
and the tunneling rate and Hawking temperature for the passed off
particles can be deduced. It is observed that the tunneling effects
are not associated to the mass of BHs but depends on the mass/energy
of the outgoing particles. In this paper we have extended the work
of charged vector particles tunneling for 5D spacetimes with
electromagnetic background. This paper is determined as follows: We
discuss in the section \textbf{2}, the tunneling rate and Hawking
temperature of charged vector $W^{\pm}$-bosons for black rings.
Section \textbf{3} is based on the analysis for Myers-Perry metric.
Section \textbf{4} is explained the outlook.

\section{Rotating Charged Black Ring}

In this section, we analyze Hawking radiation process as tunneling
of charged vector particles from 5D charged black rings. The
discussion of tunneling phenomenon for Dirac particles from 5D
charged black rings is studied in Ref.\cite{R23}. The black rings
can go around in the azimuthal focus of the $S^{2}$. The result is
identifying a revolving black ring, which takes essential conic
singularity, because at that place no centrifugal force exists to
produce equilibrium in self-gravity of black rings. Revolving black
rings in Taub-NUT have significance in Kaluza-Klein theory. The
black ring reduces to the electrically charged BH, as the NUT
parameter reduces to the similar Kaluza-Klein monopole \cite{R3}.

The study of BH thermodynamics has significance in gravitational
physics. Laws of BH thermodynamics have suggested that BHs have
finite temperature (known as Hawking temperature) which is
proportional to their surface gravity and BH entropy is proportional
to the horizon area, which corresponds to the first law of
thermodynamics. The thermodynamical relationships for 5D physical
objects at horizons are more complicated as compare to BHs, in
particular for spinning black rings and black saturns. Matsmoto
\emph{et al.} \cite{R4} have studied the time evolution of thin
black ring via Hawking radiation. The black ring is considered to
emit massless scalar particles. They observed the energy and angular
momentum of emitted scalar particles. The minimum (or zero) velocity
is deduced for black ring which allows a limit for jump particles to
be bounded inside the black ring. The analysis is based on geodesics
to analyze the complicated mathematical structure for particles to
go across through the ring \cite{R24}.

In this section, we focus on studying Hawking radiation of charged
vector particles via tunneling from 5D rotating charged black ring,
which is special solution of the Einstein-Maxwell-Dilaton gravity
model (EMD) in 5D. The line element of black rings in a unit
electric charged can be written in the following form \cite{R23}
\begin{eqnarray}
ds^{2}&&=-\frac{-F(y)}{F(x)K^{2}(x,y)}\left(dt-C(\nu,\lambda)
R\frac{1+y}{F(y)}\cosh^{2}\alpha d\phi\right)^{2} \nonumber\\
&&+\frac{R^{2}}{(x-y)^{2}}F(x)\left[-\frac{G(y)}{F(y)}d\phi^{2}
-\frac{dy^{2}}{G(y)}+\frac{dx^{2}}{G(x)}+\frac{G(x)}{F(x)}
d\varphi^{2}\right],\label{1}
\end{eqnarray}
where
\begin{eqnarray}
C(\nu,\lambda)&=&\sqrt{\lambda(\lambda-\nu)\frac{1+\lambda}{1-\lambda}}~~~~
K(x,y)=1+\lambda(x-y) \sinh^{2} \alpha \diagup F(x)\nonumber\\
F(\zeta)&=&1+\lambda\zeta,~~~~~G(\zeta)=(1-\zeta^{2})(1+\nu\zeta)\nonumber
\end{eqnarray}
and parameters $\lambda$ and $\nu$ are dimensionless and takes
values in the range of $0<\nu \leq \lambda<1$ and not takes the
conical singularity at $x=1$, $\lambda$ and $\nu$ associated to each
other say $\lambda=2\nu/(1+\nu^{2})$ and $\alpha$ is the parameter
acting as the electric charge. The coordinate $\varphi$ and $\phi$
are two cycles of the black ring and $x$ and $y$ the values rang
$-1\leq x\leq 1$ and $-\infty\leq y\leq -1$ and R has the
dimensional length. The explicit calculation of the electric charged
is $Q=2M\sinh2\alpha \diagup (3(1+\frac{4}{3}\sinh^{2}\alpha))$. The
mass of the black ring is $M = 3\pi R^2\lambda/[4(1 - \nu)]$, and
its angular momentum takes the form $J =\pi R^3p
\sqrt{\lambda(\lambda - \nu)(1 + \lambda)}/[2(1 - \nu)^2]$.

The line element given by Eq.(\ref{1}) can be rewritten as
\begin{equation}
ds^{2}=Adt^{2}+Bd\phi^{2}
+Cdy^{2}+Ddx^{2}+Ed\varphi^{2}+2Fdtd\phi
\end{equation}
The charged bosons $(W^{\pm})$ act differently as from the uncharged
bosons $(Z)$ in the presence of electromagnetic field. The $W^{+}$
bosons behave similarly as $W^{-}$ bosons and their tunneling
processes are similar too. In the background of electromagnetic
field, for anti-symmetric tensor $\psi^{\mu\nu}$, the Lagrangian
equation implies
\begin{equation}
\frac{1}{\sqrt{-g}}\partial_{\mu}(\sqrt{-g}\psi^{\nu\mu})+
\frac{m^{2}}{h^{2}}\psi^{\nu}+\frac{i}{h}e A_{\mu}\psi^{\nu\mu}+
\frac{i}{h}eF^{\nu\mu}\psi_{\mu}=0,\label{r}
\end{equation}
where
\begin{equation}
\psi_{\nu\mu}=\partial_{\nu}\psi_{\mu}-\partial_{\mu}\psi_{\nu}+
\frac{i}{h}e A_{\nu}\psi_{\mu}-\frac{i}{h}e A_{\mu}\psi_{\nu}~~
\textmd{and}~~
F^{\mu\nu}=\nabla^{\mu}A^{\nu}-\nabla^{\nu}A^{\mu}.\nonumber
\end{equation}
The values of the components of $\psi^{\mu}$ and $\psi^{\nu\mu}$ are
given as follows
\begin{eqnarray}
&&\psi^{0}=\frac{B}{AB-F^{2}}\psi_{0}-\frac{F}{AB-F^{2}}\psi_{1},~~~ \psi^{1}
=\frac{-F}{AB-F^{2}}\psi_{0}+\frac{A}{AB-F^{2}}\psi_{1},\nonumber\\
&&\psi^{2}=\frac{1}{C}\psi_{2},~~~~
\psi^{3}=\frac{1}{D}\psi_{3},~~~~
\psi^{4}=\frac{1}{E}\psi_{4},~~~\psi^{01}=\frac{F^{2}\psi_{10}+AB\psi_{01}}
{(AB-F^{2})^{2}},\nonumber\\
&&\psi^{02}=\frac{B\psi_{02}-F\psi_{12}}{C(AB-F^{2})},~~~~
\psi^{03}=\frac{B\psi_{03}-F\psi_{13}}{D(AB-F^{2})},~~~~\psi^{04}=
\frac{B\psi_{04}-F\psi_{14}}{E(AB-F^{2})},\nonumber\\
&& \psi^{12}=\frac{-F\psi_{02}+A\psi_{12}}{C(AB-F^{2})},~~
\psi^{13}=\frac{-F\psi_{03}+A\psi_{13}}{D(AB-F^{2})},
~~\psi^{14}=\frac{-F\psi_{04}+A\psi_{14}}{E(AB-F^{2})},\nonumber\\
&&\psi^{23}=\frac{\psi_{23}}{DC},~~~~~\psi^{24}=\frac{\psi_{24}}{EC},
~~~\psi^{34}=\frac{\psi_{34}}{DE}.\nonumber
\end{eqnarray}
The electromagnetic vector potential is given by
\begin{equation}
A_{\mu}=A_{t}dt+A_{\phi}d\phi,
\end{equation}
where
\begin{equation}
A_{t}=\frac{\lambda(x-y)\sinh\alpha\cosh\alpha}{F(x)K(x,y)},~~~A_{\phi}=
\frac{C(\nu,\lambda)R(1+y)\sinh\alpha\cosh\alpha}{F(x)K(x,y)}.\nonumber
\end{equation}
Using, WKB approximation \cite{R24}, i.e.,
\begin{equation}
\psi_{\nu}=c_{\nu}\exp[\frac{i}{\hbar}S_{0}(t,r,\theta,\phi)+
\Sigma \hbar^{n}S_{n}(t,r,\theta,\phi)],
\end{equation}
to the Lagrangian (\ref{r}), where $c_{\nu}$ are arbitrary
constants, $S_{0}$ and $S_{n}$ are arbitrary functions. By ignoring
the terms for $n=1,2,3,4,...$, we obtain the following set of
equations
\begin{eqnarray}
&&\frac{F^{2}}{A B-F^{2}}[C_{0}(\partial_{1}S_{0})^{2}-
C_{1}(\partial_{1}S_{0}) (\partial_{0}S_{0})+
C_{0}(\partial_{1}S_{0})eA_{1}-C_{1}(\partial_{1}S_{0})e A_{0}]\nonumber\\
&&+\frac{AB}{A B-F^{2}}[C_{1}(\partial_{1}S_{0})(\partial_{0}S_{0})-C_{0}(\partial_{1}S_{0})
+C_{1}(\partial_{1}S_{0})e A_{0}-C_{0}(\partial_{1}S_{0})e A_{1}]\nonumber\\
&&+\frac{B}{C}[C_{2}(\partial_{0}S_{0})(\partial_{2}S_{0})-C_{0}(\partial_{2}S_{0})^{2}+
C_{2}(\partial_{2}S_{0})e A_{0}]-\frac{F}{C}[C_{2}(\partial_{1}S_{0})(\partial_{2}S_{0})\nonumber\\
&&-C_{1}(\partial_{2}S_{0})^{2}+C_{2}(\partial_{2}S_{0})eA_{1}]+
\frac{B}{D}[C_{3}(\partial_{0}S_{0})(\partial_{3}S_{0})
-C_{0}(\partial_{3}S_{0})^{2}+C_{3}(\partial_{3}S_{0})\nonumber
\end{eqnarray}
\begin{eqnarray}
&& eA_{0}]-\frac{F}{D}[C_{3}(\partial_{1}S_{0})(\partial_{3}S_{0})-
C_{1}(\partial_{3}S_{0})^{2}+C_{3}(\partial_{3}S_{0})eA_{1}]
+\frac{B}{E}[C_{4}(\partial_{0}S_{0})\nonumber\\
&&(\partial_{4}S_{0})-C_{0}(\partial_{4}S_{0})^{2}+C_{4}(\partial_{4}S_{0})eA_{0}]
-\frac{F}{E}[C_{4}(\partial_{1}S_{0})(\partial_{4}S_{0})-C_{1}(\partial_{4}S_{0})^{2}\nonumber\\
&&+C_{4}(\partial_{4}S_{0})eA_{1}]-m^{2}BC_{0}+m^{2}FC_{1}
+\frac{eA_{1}F^{2}}{AB-F^{2}}[C_{0}(\partial_{1}S_{0})
-C_{1}(\partial_{0}S_{0})\nonumber\\
&&+eA_{1}C_{0}-eA_{0}C_{1}]
+\frac{eA_{1}AB}{AB-F^{2}}[C_{1}(\partial_{0}S_{0})-
C_{0}(\partial_{1}S_{0})+eA_{0}C_{1}-eA_{1}C_{0}]\nonumber\\
&&=0,\label{2}\\
&&\frac{-F^{2}}{AB-F^{2}}[C_{0}(\partial_{1}S_{0})(\partial_{0}S_{0})
-C_{1}(\partial_{0}S_{0})^{2}+C_{0}(\partial_{0}S_{0})eA_{1}-C_{1}(\partial_{0}S_{0})e A_{0}]\nonumber\\
&&-\frac{AB}{A B-F^{2}}[C_{1}(\partial_{0}S_{0})^{2}-C_{0}(\partial_{1}S_{0})
(\partial_{0}S_{0})+C_{1}(\partial_{0}S_{0})e A_{0}-C_{0}(\partial_{0}S_{0})e A_{1}]\nonumber\\
&&-\frac{F}{C}[C_{2}(\partial_{0}S_{0})(\partial_{2}S_{0})-C_{0}(\partial_{2}S_{0})^{2}+
C_{2}(\partial_{2}S_{0})e A_{0}]+\frac{A}{C}[C_{2}(\partial_{1}S_{0})(\partial_{2}S_{0})\nonumber\\
&&-C_{1}(\partial_{2}S_{0})^{2}+C_{2}(\partial_{2}S_{0})eA_{1}]
-\frac{F}{D}[C_{3}(\partial_{0}S_{0})(\partial_{3}S_{0})
-C_{0}(\partial_{3}S_{0})^{2}+C_{3}(\partial_{3}S_{0})\nonumber\\
&&eA_{0}]+\frac{A}{D}[C_{3}(\partial_{1}S_{0})(\partial_{3}S_{0})-
C_{1}(\partial_{3}S_{0})^{2}+C_{3}(\partial_{3}S_{0})eA_{1}]
-\frac{F}{E}[C_{4}(\partial_{0}S_{0})\nonumber\\
&&(\partial_{4}S_{0})-C_{0}(\partial_{4}S_{0})^{2}+C_{4}(\partial_{4}S_{0})eA_{0}]
+\frac{A}{E}[C_{4}(\partial_{1}S_{0})(\partial_{4}S_{0})-C_{1}(\partial_{4}S_{0})^{2}
\nonumber\\&&+C_{4}(\partial_{4}S_{0})eA_{1}]+m^{2}FC_{0}-m^{2}AC_{1}
-\frac{eA_{0}F^{2}}{AB-F^{2}}[C_{0}(\partial_{1}S_{0})
-C_{1}(\partial_{0}S_{0})\nonumber\\&&-eA_{1}C_{0}+eA_{0}C_{1}]
+\frac{eA_{0}AB}{AB-F^{2}}[C_{3}(\partial_{1}S_{0})-
C_{1}(\partial_{3}S_{0})+eA_{1}C_{3}]=0,\\
&&\frac{F}{A B-F^{2}}[C_{2}(\partial_{1}S_{0})(\partial_{0}S_{0})-
C_{1}(\partial_{0}S_{0})(\partial_{1}S_{0})+C_{2}(\partial_{0}S_{0})eA_{1}]\nonumber\\
&&-\frac{B}{A B-F^{2}}[C_{2}(\partial_{0}S_{0})^{2}-C_{0}(\partial_{0}S_{0})
(\partial_{2}S_{0})+C_{2}(\partial_{0}S_{0})e A_{0}]\nonumber\\
&&+\frac{F}{AB-F^{2}}[C_{2}(\partial_{0}S_{1})(\partial_{0}S_{0})
-C_{0}(\partial_{2}S_{0})(\partial_{1}S_{0})+C_{2}(\partial_{1}S_{0})e A_{0}]\nonumber\\
&&-\frac{A}{AB-F^{2}}[C_{2}(\partial_{1}S_{0})^{2}-C_{1}(\partial_{2}S_{0})(\partial_{1}S_{0})
+C_{2}(\partial_{1}S_{0})eA_{1}]-\frac{1}{D}[C_{3}(\partial_{2}S_{0})\nonumber\\
&&(\partial_{3}S_{0})-C_{2}(\partial_{3}S_{0})^{2}]
+\frac{1}{E}[C_{4}(\partial_{4}S_{0})(\partial_{2}S_{0})-
C_{2}(\partial_{4}S_{0})^{2}]-m^{2}C_{2}+\nonumber\\
&&\frac{eA_{0}F}{AB-F^{2}}[C_{2}(\partial_{1}S_{0})-C_{1}(\partial_{2}S_{0})+C_{2}eA_{1}]
-\frac{B}{AB-F^{2}}[C_{2}(\partial_{0}S_{0})\nonumber
\end{eqnarray}
\begin{eqnarray}
&&-C_{0}(\partial_{2}S_{0})+C_{2}eA_{0}]
+\frac{eA_{1}F}{AB-F^{2}}[C_{2}(\partial_{0}S_{0})
-C_{0}(\partial_{2}S_{0})+eA_{0}C_{2}]\nonumber\\
&&-\frac{eA_{1}A}{AB-F^{2}}[C_{2}(\partial_{1}S_{0})-
C_{1}(\partial_{2}S_{0})+eA_{1}C_{2}]=0,\\
&&-B[C_{3}(\partial_{0}S_{0})^{2}-C_{0}(\partial_{0}S_{0})(\partial_{3}S_{0})+
C_{3}(\partial_{0}S_{0})eA_{0}]+F[C_{3}(\partial_{0}S_{0})(\partial_{1}S_{0})\nonumber\\
&&-C_{1}(\partial_{0}S_{0})(\partial_{3}S_{0})+C_{3}(\partial_{0}S_{0})e
A_{1}]+F[C_{3}(\partial_{0}S_{0})(\partial_{1}S_{0})-C_{0}(\partial_{1}S_{0})(\partial_{3}S_{0})\nonumber\\
&&+C_{3}(\partial_{1}S_{0})e A_{0}]-A[C_{3}(\partial_{1}S_{0})^{2}
-C_{1}(\partial_{1}S_{0})(\partial_{3}S_{0})
+C_{3}(\partial_{1}S_{0})eA_{1}]\nonumber\\
&&-\frac{AB-F^{2}}{C}[C_{3}(\partial_{2}S_{0})^{2}
-C_{2}(\partial_{2}S_{0})(\partial_{3}S_{0})]+\frac{AB-F^{2}}
{E}[C_{4}(\partial_{3}S_{0})(\partial_{4}S_{0})-\nonumber\\
&&C_{3}(\partial_{4}S_{0})^{2}]-m^{2}C_{3}\frac{AB-F^{2}}{D}+eA_{0}B[C_{3}(\partial_{0}S_{0})
-C_{0}(\partial_{3}S_{0})+C_{3}eA_{0}]\nonumber\\
&&+eA_{0}F[C_{3}(\partial_{1}S_{0})-C_{1}(\partial_{3}S_{0})
+C_{3}eA_{1}]+eA_{1}F[C_{3}(\partial_{0}S_{0})
-C_{0}(\partial_{3}S_{0})+\nonumber\\
&&eA_{0}C_{3}]-eA_{1}A[C_{3}(\partial_{1}S_{0})-
C_{1}(\partial_{3}S_{0})+eA_{1}C_{3}]=0,\\
&&F[C_{4}(\partial_{0}S_{0})(\partial_{1}S_{0})-C_{1}
(\partial_{0}S_{0})(\partial_{4}S_{0})+
C_{4}(\partial_{0}S_{0})eA_{1}]-B[C_{4}(\partial_{0}S_{0})^{2}\nonumber\\
&&-C_{0}(\partial_{0}S_{0})(\partial_{4}S_{0})+C_{4}(\partial_{0}S_{0})e
A_{0}] +F[C_{4}(\partial_{0}S_{0})(\partial_{1}S_{0})-
C_{0}(\partial_{1}S_{0})(\partial_{4}S_{0})\nonumber\\
&&+C_{4}(\partial_{1}S_{0})e A_{0}]-A[C_{4}(\partial_{1}S_{0})^{2}
-C_{1}(\partial_{1}S_{0})(\partial_{4}S_{0})
+C_{4}(\partial_{1}S_{0})eA_{1}]\nonumber\\
&&-\frac{AB-F^{2}}{C}[C_{4}(\partial_{2}S_{0})^{2}
-C_{2}(\partial_{2}S_{0})(\partial_{4}S_{0})]
-\frac{AB-F^{2}}{D}[C_{4}(\partial_{3}S_{0})^{2}\nonumber\\
&&-C_{3}(\partial_{4}S_{0})(\partial_{3}S_{0})]
-m^{2}C_{4}AB-F^{2}+eA_{0}F[C_{4}(\partial_{1}S_{0})
-C_{1}(\partial_{4}S_{0})\nonumber\\
&&+C_{4}eA_{1}]-eA_{0}B[C_{4}(\partial_{0}S_{0})-C_{0}(\partial_{4}S_{0})
+C_{4}eA_{0}]+eA_{1}F[C_{4}(\partial_{0}S_{0})\nonumber\\
&&-C_{0}(\partial_{4}S_{0})+eA_{0}C_{4}]
-eA_{1}A[C_{4}(\partial_{1}S_{0})-C_{1}(\partial_{4}S_{0})+eA_{1}C_{4}]=0.\label{3}
\end{eqnarray}
Using the following rule for separation of variables \cite{R25},
i.e.,
\begin{equation}
S_{0}=-(E-\sum_{i=1}^{2}j_{i}\check{\Omega}_{i})t+J\phi+W(x,y)+L\varphi,\label{RR1}
\end{equation}
where $E$ is the energy of the particle, $J$ and $L$ represent the
particle's angular momentums corresponding to the angles $\phi$ and
$\varphi$, respectively, while $K$ is the complex constant. From
Eqs.(\ref{2})-(\ref{3}), we can obtain a $5\times5$ matrix equation
\begin{equation*}
G(C_{0},C_{1},C_{2},C_{3},C_{4})^{T}=0,
\end{equation*}
where the matrix is labeled as $``\textbf{G}"$, whose components are
given as follows:
\begin{eqnarray}
G_{00}&=&\frac{1}{AB-F^{2}}[F^{2}J^{2}+eA_{1}]-
AB[J-eA_{1}J]-\frac{1}{C}(\dot{W}^{2}-\frac{1}{D}(\partial_{3}W)^{2}\nonumber\\&-&\frac{L^{2}}{E}
-m^{2}B +\frac{eA_{1}}{AB-F^{2}}[F^{2}J-eA_{1}]-AB[J+eA_{1}]\nonumber\\
G_{01}&=&\frac{1}{AB-F^{2}}[(E-\sum_{i=1}^{2}j_{i}\check{\Omega}_{i})J-eA_{0}J]
-AB[(E-\sum_{i=1}^{2}j_{i}\check{\Omega}_{i})J-eA_{0}J]\nonumber\\
&+&F[(\partial_{2}W)^{2}+eA_{2}(\partial_{2}W)]-F[(\partial_{3}W)^{2}-L^{2}]+\frac{eA_{1}}{AB-F^{2}}
[(E-\sum_{i=1}^{2}j_{i}\check{\Omega}_{i})\nonumber\\
&-&eA_{0}]-AB[(E-\sum_{i=1}^{2}j_{i}\check{\Omega}_{i})-eA_{0}]\nonumber\\
G_{02}&=&\frac{B}{C}[(E-\sum_{i=1}^{2}j_{i}\check{\Omega}_{i})(\partial_{2}W)+eA_{0}(\partial_{2}W)]
-F[J(\partial_{3}W)+eA_{1}(\partial_{2}W)]\nonumber\\
G_{03}&=&\frac{B}{D}[(E-\sum_{i=1}^{2}j_{i}\check{\Omega}_{i})(\partial_{3}W)+eA_{0}(\partial_{3}W)]
-F[J(\partial_{3}W)+eA_{1}(\partial_{3}W)]\nonumber\\
G_{04}&=&\frac{B}{E}[(E-\sum_{i=1}^{2}j_{i}\check{\Omega}_{i})L+eA_{0}L]
+F[(E-\sum_{i=1}^{2}j_{i}\check{\Omega}_{i})L-eA_{1}L]\nonumber\\
G_{10}&=&\frac{F^{2}}{AB-F^{2}}[(E-\sum_{i=1}^{2}j_{i}\check{\Omega}_{i})J
+eA_{1}(E-j\check{\Omega})]+AB[(E-\sum_{i=1}^{2}j_{i}\check{\Omega}_{i})J\nonumber\\
&&-eA_{1}(E-\sum_{i=1}^{2}j_{i}\check{\Omega}_{i})]+\frac{F}{C}
(\partial_{2}W)^{2}+\frac{F}{D}(\partial_{3}W)^{2}+\frac{F}{E}L^{2}+m^{2}F\nonumber\\
&&-\frac{eA_{0}F^{2}}{AB-F^{2}}[J-eA_{0}]\nonumber\\
G_{11}&=&\frac{1}{AB-F^{2}}[(E-\sum_{i=1}^{2}j_{i}\check{\Omega}_{i})^{2}
-eA_{0}(E-\sum_{i=1}^{2}j_{i}\check{\Omega}_{i})]+
AB[(E-\sum_{i=1}^{2}j_{i}\check{\Omega}_{i})^{2}\nonumber\\
&&-eA_{0}(E-\sum_{i=1}^{2}j_{i}\check{\Omega}_{i})]-\frac{A}{C}
(\partial_{2}W)+\frac{1}{D}(\partial_{3}W)^{2}A-\frac{1}{E}AL^{2}-m^{2}A\nonumber\\
&&-\frac{eA_{0}F^{2}}{AB-F^{2}}[(E-\sum_{i=1}^{2}j_{i}
\check{\Omega}_{i})-eA_{0}-(\partial_{3}W)AB]\nonumber
\end{eqnarray}
\begin{eqnarray}
G_{12}&=&\frac{F}{C}[(E-\sum_{i=1}^{2}j_{i}\check{\Omega}_{i})+eA_{0}(\partial_{2}W)]
+\frac{A}{C}[J(\partial_{2}W)+eA_{1}(\partial_{2}W)]\nonumber\\
G_{13}&=&\frac{F}{D}[(E-\sum_{i=1}^{2}j_{i}\check{\Omega}_{i})
(\partial_{3}W)-eA_{0}(\partial_{3}W)]+\frac{A}{D}[J(\partial_{3}W)+eA_{1}(\partial_{3}W)]\nonumber\\
&+&\frac{eA_{0}}{AB-F^{2}}[ABJ+ABeA_{1}]\nonumber\\
G_{14}&=&\frac{F}{E}[(E-\sum_{i=1}^{2}j_{i}\check{\Omega}_{i})L-
eA_{0}L]+\frac{A}{E}[JL+eA_{1}L]\nonumber\\
G_{20}&=&\frac{-B}{AB-F^{2}}(E-\sum_{i=1}^{2}j_{i}\check{\Omega}_{i})
(\partial_{2}W)-\frac{1}{AB-F^{2}}(\partial_{2}W)
+\frac{eA_{0}B}{AB-F^{2}}(\partial_{2}W)\nonumber\\
&-&\frac{eA_{1}}{AB-F^{2}}(\partial_{2}W)\nonumber\\
G_{21}&=&\frac{F}{AB-F^{2}}(E-\sum_{i=1}^{2}j_{i}\check{\Omega}_{i})
(\partial_{2}W)+\frac{A}{AB-F^{2}}J(\partial_{2}W)
-\frac{eA_{0}}{AB-F^{2}}(\partial_{2}W)\nonumber\\
&+&\frac{eA_{1}A}{AB-F^{2}}(\partial_{2}W)\nonumber\\
G_{22}&=&\frac{-F}{AB-F^{2}}[J(E-\sum_{i=1}^{2}j_{i}\check{\Omega}_{i})
+eA_{1}(E-\sum_{i=1}^{2}j_{i}\check{\Omega}_{i})]-\frac{B}{AB-F^{2}}[(E\nonumber\\
&&-\sum_{i=1}^{2}j_{i}\check{\Omega}_{i})^{2}-eA_{0}(E-\sum_{i=1}^{2}j_{i}\check{\Omega}_{i})]
-\frac{F}{AB-F^{2}}[(E-\sum_{i=1}^{2}j_{i}\check{\Omega}_{i})J-eA_{0}J]\nonumber\\
&&-\frac{A}{AB-F^{2}}[J^{2}+eA_{1}J]-\frac{1}{D}(\partial_{3}W)^{2}
-\frac{L^{2}}{E}-m^{2}+\frac{eA_{0}F}{AB-F^{2}}[J+eA_{1}]\nonumber\\
&+&\frac{B}{AB-F^{2}}[(E-\sum_{i=1}^{2}j_{i}\check{\Omega}_{i})+eA_{0}]-
\frac{eA_{1}F}{AB-F^{2}}[(E-\sum_{i=1}^{2}j_{i}\check{\Omega}_{i})+eA_{0}]\nonumber\\
&-&\frac{eA_{1}A}{AB-F^{2}}[J+eA_{1}],~~~~G_{23}=\frac{1}{D}
(\partial_{2}W)(\partial_{3}W)~~~~G_{24}=\frac{1}{E}(\partial_{2}W)L\nonumber\\
G_{30}&=&-B(\partial_{3}W)(E-\sum_{i=1}^{2}j_{i}\check{\Omega}_{i})-
FJ(\partial_{3}W)-eA_{0}(\partial_{3}W)+eA_{1}F(\partial_{3}W)\nonumber\\
G_{31}&=&F(E-\sum_{i=1}^{2}j_{i}\check{\Omega}_{i})(\partial_{3}W)
+AJ-eA_{0}F(\partial_{3}W)+eA_{1}A(\partial_{3}W)\nonumber\\
G_{32}&=&\frac{AB-F^{2}}{C}(\partial_{2}W)(\partial_{3}W)\nonumber
\end{eqnarray}
\begin{eqnarray}
G_{33}&=&-B[(E-\sum_{i=1}^{2}j_{i}\check{\Omega}_{i})^{2}-eA_{0}(E-j\check{\Omega})]
-F[(E-\sum_{i=1}^{2}j_{i}\check{\Omega}_{i})J\nonumber\\
&&-eA_{1}(E-\sum_{i=1}^{2}j_{i}\check{\Omega}_{i})]-F[(E-
\sum_{i=1}^{2}j_{i}\check{\Omega}_{i})J-eA_{0}J]-\nonumber\\
&&A[J^{2}+eA_{1}J]-\frac{AB-F^{2}}{C}(\partial_{2}W)^{2}-\frac{AB-F^{2}}{E}L-\nonumber\\
&&\frac{AB-F^{2}}{D}m^{2}-eA_{0}B[(E-\sum_{i=1}^{2}j_{i}\check{\Omega}_{i})-eA_{0}]+eA_{0}F[J+eA_{1}]\nonumber\\
&&-eA_{1}F[(E-\sum_{i=1}^{2}j_{i}\check{\Omega}_{i})-eA_{0}]-AeA_{1}[J+eA_{1}]\nonumber\\
G_{34}&=&\frac{AB-F^{2}}{E}(\partial_{3}W)L\nonumber\\
G_{40}&=&-B(E-\sum_{i=1}^{2}j_{i}\check{\Omega}_{i})L-FJL+eA_{0}BL-eA_{1}FL\nonumber\\
G_{41}&=&F(E-\sum_{i=1}^{2}j_{i}\check{\Omega}_{i})L+AJL-eA_{0}FL+eA_{1}A\nonumber\\
G_{42}&=&\frac{AB-F^{2}}{C}(\partial_{2}W)L,~~~
G_{43}=\frac{AB-F^{2}}{D}(\partial_{3}W)L\nonumber\\
G_{44}&=&-F(E-\sum_{i=1}^{2}j_{i}\check{\Omega}_{i})J-FeA_{1}(E-j\check{\Omega})
-B[(E-\sum_{i=1}^{2}j_{i}\check{\Omega}_{i})^{2}-\nonumber\\
&&eA_{0}(E-\sum_{i=1}^{2}j_{i}\check{\Omega}_{i})]
-F[(E-\sum_{i=1}^{2}j_{i}\check{\Omega}_{i})-eA_{0}J]-A[J^{2}+eA_{1}J]\nonumber\\
&&-\frac{AB-F^{2}}{C}(\partial_{2}W)^{2}-\frac{AB-F^{2}}{D}(\partial_{3}W)^{2}
-m^{2}(AB-F^{2})+eA_{0}F\nonumber\\
&&[J+eA_{1}]+eA_{0}B[(E-\sum_{i=1}^{2}j_{i}\check{\Omega}_{i})-eA_{0}]-eA_{1}F
[(E-\sum_{i=1}^{2}j_{i}\check{\Omega}_{i})\nonumber\\
&&+eA_{0}]-eA_{1}A[J+eA_{1}]\nonumber
\end{eqnarray}
For the non-trivial solution, the absolute value $|\textbf{G}|$ is
equal to \emph{zero}, so that one can yield
\begin{eqnarray}
ImW^{\pm}&&=\pm \int\sqrt{\frac{(E-\sum_{i=1}^{2}j_{i}\check{\Omega}_{i}
-eA_{0}-\Omega_{h}eA_{1})^{2}+X}{\frac{F^{2}-AB}{BD}}},\nonumber\\
&&=\pm i\pi\frac{E-eA_{0}-\sum_{i=1}^{2}j_{i}\check{\Omega}_{i}
-\Omega_{h}eA_{1}}{2\kappa(r_{+})},
\end{eqnarray}
where $-\Omega_{h}=\frac{F^{2}}{B^{2}}$, while $+$ and $-$ represent
the outgoing and incoming particles, respectively. The function $X$
can be defined as
\begin{eqnarray}
X&=&\frac{FJ}{B}(E-j\check{\Omega})-e^{2}A^{2}_{1}\frac{F}{B}[\frac{F}{B}-1]
+\frac{FJ}{B}[(E-j\check{\Omega})-eA_{0}]+\frac{AJ}{B}\nonumber\\&&
[J+eA_{1}] +m^{2}\frac{(AB-F^{2})}{B}-eA_{0}\frac{F}{B}J
+eA_{1}\frac{A}{B}(J+eA_{1})
\end{eqnarray}
and the surface gravity is
\begin{equation}
\kappa(r_{+})=\frac{1}{2}\sqrt{K_{y}(x,y)H_{y}(x,y)},\nonumber
\end{equation}
where we have choosed $-C_{y}(x,y)=K_{y}(x,y)~\textmd{and}
~D^{-1}_{y}(x,y)=H_{y}(x,y)$. The tunneling probability for charged
vector particles is given by
\begin{equation}
\Gamma=\frac{\Gamma_{(emission)}}{\Gamma_{(absorption)}}
=\exp\left[-4\pi \frac{(E-\sum_{i=1}^{2}j_{i}\check{\Omega}_{i}
-eA_{0}-\Omega_{h}eA_{1})}{\sqrt{K_{y}(x,y)H_{y}(x,y)}}\right].\label{W3}
\end{equation}
By comparing the above expression with the Boltzmann factor,
$\exp[-\beta E]$, one can derive the Hawking temperature, which is
$T_{H}=\frac{1}{\beta}$ at the outer horizon $r_{+}$. For this case,
we can obtain the following Hawking temperature as
\begin{eqnarray}
T_{H}&&=\frac{\sqrt{K_{y}(x,y)H_{y}(x,y)}}{4\pi},\nonumber\\
&&=\frac{\sqrt{(x-y)^{2}(1-x^{2})(1+\upsilon x)[2+\upsilon
y+\upsilon
y^{3}+x\upsilon-2xy-3xy^{2}\upsilon]}}{4\pi\sqrt{(1-y^{2})(1+y\upsilon)}}.\label{W4}
\end{eqnarray}

The Hawking temperature of electrically charged black ring is
depending on $x$, $y$ and $\upsilon$. The resulting Hawking
temperature at which vector particles tunnel through the horizon is
similar to the Hawking temperature for scalar and Dirac particles at
which they tunnel through the horizon of black ring \cite{R23}.

\section{Myers-Perrry Black Hole}

Black holes are most valuable astrophysical objects predicted in
general relativity \cite{R26}. Kaluza's theory (1921) along with
Klein's version (1929) known as Kaluza-Klein theory which provides
the proposal to unify the theory of general relativity and
electromagnetic theory in 5D vacuum spacetime. The Einstein field
equations for 5D spacetime are equivalent to 4D Einstein's equation
on with matter comprising of scalar and electromagnetic fields.
Abdolrahimi et al. \cite{R27} have studied distorted Myers-Perry BH
in an external gravitational field with a single angular momentum as
exact solution of the 5D Einstein equations in vacuum.

The Myers-Perry BH \cite{W1} is a solution of vacuum Einstein field
equations in an arbitrary dimension. Here, we consider the 5D case
which represents a regular rotating BH (with two rotation
parameters) which is a generalization of the Kerr solution. The
corresponding line element in the Boyer-Lindquist coordinates $(t,
r, \theta, \phi, \varphi)$ is defined as follows \cite{R28}
\begin{eqnarray}
ds^{2}&=&\frac{\rho^{2}r^{2}}{\Delta}dr^{2}+\rho^{2}d\theta^{2}-dt^{2}
+(r^{2}+a^{2})\sin^{2}\theta d\phi^{2}+(r^{2}+b^{2})\cos^{2}\theta
d\varphi^{2}\nonumber\\
&+&\frac{r_{0}^{2}}{\rho^{2}}[dt+a\sin^{2}\theta
d\phi+b\cos^{2}\theta d\varphi]^{2}\label{aa}
\end{eqnarray}
where
\begin{equation*}
\rho^{2}=r^{2}+a^{2}\cos^{2}\theta+b^{2}\sin^{2}\theta,~~~
\Delta=(r^{2}+a^{2})(r^{2}+b^{2})-r_{0}^{2}r^{2},
\end{equation*}
and the angle $\phi$ and $\varphi$ assume measures from the interval
$[0, 2\pi]$ and $\theta$ assume measures in $[0, \frac{\pi}{2}]$;
$a$ and $b$ are the two angular momenta, $r_{0}$ associated to the
BH mass. The BH horizons are turned as
\begin{equation*}
r^{2}_{\pm}=\frac{1}{2}[r^{2}_{0}-a^{2}-b^{2}\pm\sqrt{(r^{2}_{0}-a^{2}-b^{2})^{2}-4a^{2}b^{2}}]
~~~\textmd{for}~~~(r^{2}_{0}-a^{2}-b^{2})^{2}>4a^{2}b^{2}\nonumber
\end{equation*}
The Myers-Perry BH is regular except if $a=b=0$ that corresponds to
5D Schwarzschild-Tangherlini BH solution \cite{W2} and only in this
case it has a singularity at $r=0$.

The line element given by Eq.(\ref{aa}) can be rewritten as
\begin{equation}
ds^{2}=\tilde{A} dt^{2}+\tilde{B} dr^{2}+ \tilde{C}
d\theta^{2}+\tilde{D} d\phi^{2}+\tilde{E} d\varphi^{2}+ 2\tilde{F}
dt d\phi+2\tilde{G} dt d\varphi+2\tilde{H} d\phi d\varphi.
\end{equation}
The electromagnetic vector potential is defined as
$A_{\mu}=(A_{1},0,0,A_{4},0)$, where
\begin{eqnarray*}
A_{1}=\frac{b}{\kappa\tilde{\phi^{2}}}b\cos^{2}\theta\frac{r_{0}^{2}}{\rho^{2}}~~
\textmd{and}~~
A_{4}=\frac{r_{0}^{2}}{\rho^{2}{\kappa\tilde{\phi^{2}}}}ab\sin^{2}\theta.
\cos^{2}\theta
\end{eqnarray*}
In Lagrangian Eq.(\ref{r}) the components of $\psi^{\nu}$ and
$\psi^{\mu\nu}$ are given by
\begin{eqnarray}
\psi^{0}&=&J\psi_{0}+M\psi_{3}+N\psi_{4},~~\psi^{1}=T\psi_{1},~~\psi^{2}=U\psi_{2},\nonumber\\
\psi^{3}&=&M\psi_{0}+P\psi_{3}+R\psi_{4},~~\psi^{4}=\tilde{E}^{-1}\psi_{4},\nonumber\\
\psi^{01}&=&JT\psi_{01}+MT\psi_{31}+NT\psi_{41},~~\psi^{02}=JU\psi_{02}+MU\psi_{32}+NU\psi_{42},\nonumber\\
\psi^{03}&=&(J P-M^{2})\psi_{03}+(J R-M N)\psi_{04}+(MR-NP)\psi_{34},\nonumber\\
\psi^{04}&=&MN\psi_{30}+(M^{2}-J S)\psi_{40}+JR\psi_{03}+NR\psi_{43},~~\psi^{12}=TU\psi_{12},\nonumber\\
\psi^{13}&=&TM\psi_{10}+TP\psi_{13}+T R\psi_{14},~~\psi^{14}=TN\psi_{10}+TR\psi_{13}+TS\psi_{14},\nonumber\\
\psi^{23}&=&MU\psi_{20}+UP\psi_{23}+UR\psi_{24},~~\psi^{24}=MU\psi_{20}+UP\psi_{23}+UR\psi_{24},\nonumber\\
\psi^{34}&=&(MR-PN)\psi_{03}+PS\psi_{34},\nonumber
\end{eqnarray}
where
\begin{eqnarray}
&&J=\frac{DE-H^{2}}{-ADE+AH^{2}+F^{2}E-2HFG+DG^{2}},\nonumber\\
&& M=\frac{FE-GH}{-ADE+AH^{2}+F^{2}E-2HFG+DG^{2}},\nonumber\\
&&N=\frac{FH-GD}{-ADE+AH^{2}+F^{2}E-2HFG+DG^{2}},\nonumber\\
&&P=\frac{EA-G^{2}}{-ADE+AH^{2}+F^{2}E-2HFG+DG^{2}},\nonumber\\
&&R=\frac{-HA-FG}{-ADE+AH^{2}+F^{2}E-2HFG+DG^{2}},\nonumber\\
&&S=\frac{-DE+F^{2}}{-ADE+AH^{2}+F^{2}E-2HFG+DG^{2}},\nonumber\\
&&T=\frac{1}{B},~~~U=\frac{1}{C}.\nonumber
\end{eqnarray}
Using Eq.(\ref{r}) and WKB approximation, by neglecting the terms of
order $n=1,2,3,4...$ we obtain the following set of equations
\begin{eqnarray}
&&JT[C_{1}(\partial_{0}S_{0})(\partial_{1}S_{0})-C_{0}(\partial_{1}S_{0})^{2}+
eA_{0}C_{1}(\partial_{1}S_{0})]+MT[C_{1}(\partial_{1}S_{0})(\partial_{3}S_{0})\nonumber\\
&&-C_{3}(\partial_{1}S_{0})^{2}+eA_{3}C_{1}(\partial_{1}S_{0})]+NT[C_{1}(\partial_{1}S_{0})
(\partial_{4}S_{0})-C_{4}(\partial_{1}S_{0})^{2}]+JU[C_{2}\nonumber\\
&&(\partial_{0}S_{0})(\partial_{2}S_{0})-C_{0}(\partial_{2}S_{0})^{2}+eA_{0}C_{2}(\partial_{2}S_{0})]
+MU[C_{2}(\partial_{2}S_{0})(\partial_{3}S_{0})-C_{3}\nonumber\\
&&(\partial_{2}S_{0})^{2}+eA_{3}C_{2}(\partial_{2}S_{0})]+NU[C_{2}(\partial_{2}S_{0})
(\partial_{4}S_{0})-C_{4}(\partial_{2}S_{0})^{2}]+[JP-M^{2}]\nonumber\\
&&[C_{3}(\partial_{0}S_{0})(\partial_{3}S_{0})-C_{0}(\partial_{3}S_{0})^{2}
+eA_{3}C_{2}(\partial_{2}S_{0})]+[JR-MN][C_{4}(\partial_{0}S_{0})(\partial_{3}S_{0})\nonumber\\
&&-C_{0}(\partial_{4}S_{0})(\partial_{3}S_{0})+eA_{0}C_{4}(\partial_{3}S_{0})]+[MR-NP]
[C_{4}(\partial_{3}S_{0})^{2}-C_{3}(\partial_{4}S_{0})(\partial_{3}S_{0})\nonumber\\
&&+eA_{3}C_{4}(\partial_{3}S_{0})]+[JR-MN][C_{3}(\partial_{0}S_{0})
(\partial_{4}S_{0})-C_{0}(\partial_{3}S_{0})(\partial_{4}S_{0})+C_{3}(\partial_{4}S_{0})\nonumber\\
&&-eA_{3}C_{0}(\partial_{4}S_{0})]+[JS-M^{2}][C_{4}(\partial_{0}S_{0})(\partial_{4}S_{0})
-C_{0}(\partial_{4}S_{0})^{2}+eA_{0}C_{4}(\partial_{4}S_{0})]\nonumber\\
&&+NR[C_{3}(\partial_{4}S_{0})^{2}-C_{4}(\partial_{3}S_{0})(\partial_{4}S_{0})
-eA_{3}C_{4}(\partial_{4}S_{0})]-m^{2}[C_{0}J+C_{3}M+\nonumber\\
&&C_{4}N]+eA_{3}[C_{3}(\partial_{0}S_{0})-C_{0}(\partial_{3}S_{0})+eA_{0}C_{3}
-eA_{3}C_{0}]+eA_{3}[JR-MN]\nonumber\\
&&[C_{4}(\partial_{0}S_{0})-C_{0}(\partial_{4}S_{0})+eA_{0}C_{4}]
+eA_{3}[MR-NP][C_{4}(\partial_{3}S_{0})-C_{3}(\partial_{4}S_{0})\nonumber\\
&&+eA_{3}C_{4}]=0\label{t1}\\
&&-JT[C_{1}(\partial_{0}S_{0})^{2}-C_{0}(\partial_{0}S_{0})(\partial_{1}S_{0})+
eA_{0}C_{1}(\partial_{0}S_{0})]-MT[C_{1}(\partial_{0}S_{0})(\partial_{3}S_{0})\nonumber\\
&&-C_{3}(\partial_{0}S_{0})(\partial_{1}S_{0})+eA_{3}C_{1}(\partial_{0}S_{0})]
-NT[C_{1}(\partial_{0}S_{0})(\partial_{4}S_{0})-C_{4}(\partial_{0}S_{0})(\partial_{1}S_{0})]\nonumber\\
&&+TU[C_{2}(\partial_{1}S_{0})(\partial_{2}S_{0})-C_{1}(\partial_{2}S_{0})^{2}]+TM[C_{0}
(\partial_{1}S_{0})(\partial_{3}S_{0})-C_{1}(\partial_{0}S_{0})(\partial_{3}S_{0})\nonumber\\
&&-eA_{0}C_{1}(\partial_{3}S_{0})]+TP[C_{3}(\partial_{1}S_{0})(\partial_{3}S_{0})
-C_{1}(\partial_{3}S_{0})^{2}-eA_{3}C_{1}(\partial_{3}S_{0})]+TR\nonumber\\
&&[C_{4}(\partial_{1}S_{0})(\partial_{3}S_{0})-C_{1}(\partial_{3}S_{0})(\partial_{4}S_{0})]
+TN[C_{0}(\partial_{1}S_{0})(\partial_{4}S_{0})-C_{1}(\partial_{0}S_{0})(\partial_{4}S_{0})\nonumber\\
&&-eA_{0}C_{1}(\partial_{4}S_{0})]+TR[C_{3}(\partial_{1}S_{0})(\partial_{4}S_{0})
-C_{1}(\partial_{3}S_{0})(\partial_{4}S_{0})-eA_{3}C_{1}(\partial_{4}S_{0})]\nonumber\\
&&+TS[C_{4}(\partial_{1}S_{0})(\partial_{4}S_{0})-C_{1}(\partial_{4}S_{0})^{2}]
-m^{2}TC_{1}-eA_{0}JT[C_{0}(\partial_{1}S_{0})-C_{1}(\partial_{0}S_{0})^{2}\nonumber\\
&&-eA_{0}C_{1}]-MTeA_{0}[C_{1}(\partial_{3}S_{0})-C_{3}(\partial_{1}S_{0})
-eA_{3}C_{1}]-eA_{0}NT[C_{1}(\partial_{4}S_{0})-C_{4}\nonumber\\
&&(\partial_{1}S_{0})]+TMeA_{3}[C_{0}(\partial_{1}S_{0})-C_{1}(\partial_{0}S_{0})
-eA_{0}C_{1}]+eA_{3}TP[C_{3}(\partial_{1}S_{0})-C_{1}\nonumber\\
&&(\partial_{3}S_{0})-eA_{3}C_{1}]+TReA_{3}[C_{4}(\partial_{1}S_{0})-C_{1}(\partial_{4}S_{0})]=0,\label{t2}\\
&&JU[C_{0}(\partial_{0}S_{0})(\partial_{2}S_{0})-C_{2}(\partial_{0}S_{0})^{2}-
eA_{0}C_{2}(\partial_{0}S_{0})]-MU[C_{2}(\partial_{0}S_{0})(\partial_{3}S_{0})\nonumber\\
&&-C_{3}(\partial_{0}S_{0})(\partial_{2}S_{0})+eA_{3}C_{2}(\partial_{0}S_{0})]
-NU[C_{2}(\partial_{0}S_{0})(\partial_{4}S_{0})-C_{4}(\partial_{0}S_{0})(\partial_{2}S_{0})]\nonumber\\
&&-TU[C_{2}(\partial_{1}S_{0})^{2}-C_{1}(\partial_{2}S_{0})(\partial_{1}S_{0})]+MU[C_{0}
(\partial_{2}S_{0})(\partial_{3}S_{0})-C_{2}(\partial_{0}S_{0})(\partial_{3}S_{0})\nonumber\\
&&-eA_{0}C_{2}(\partial_{3}S_{0})]+UP[C_{3}(\partial_{2}S_{0})(\partial_{3}S_{0})
-C_{2}(\partial_{3}S_{0})^{2}-eA_{3}C_{2}(\partial_{3}S_{0})]+UR\nonumber\\
&&[C_{4}(\partial_{2}S_{0})(\partial_{3}S_{0})-C_{2}(\partial_{3}S_{0})(\partial_{4}S_{0})]
+MN[C_{0}(\partial_{2}S_{0})(\partial_{4}S_{0})-C_{2}(\partial_{0}S_{0})(\partial_{4}S_{0})\nonumber\\
&&-eA_{0}C_{2}(\partial_{4}S_{0})]+UP[C_{3}(\partial_{4}S_{0})(\partial_{2}S_{0})-C_{2}
(\partial_{3}S_{0})(\partial_{4}S_{0})]+UR[C_{4}(\partial_{4}S_{0})\nonumber\\
&&(\partial_{2}S_{0})-C_{2}(\partial_{4}S_{0})^{2}]-m^{2}UC_{2}-eA_{0}JU
[C_{0}(\partial_{2}S_{0})-C_{2}(\partial_{0}S_{0})-eA_{0}C_{2}]\nonumber\\
&&-MUeA_{0}[C_{2}(\partial_{3}S_{0})-C_{3}(\partial_{2}S_{0})]
-eA_{0}NU[C_{2}(\partial_{4}S_{0})-C_{4}(\partial_{2}S_{0})]+UM\nonumber\\
&&eA_{3}[C_{0}(\partial_{2}S_{0})-C_{2}(\partial_{0}S_{0})-eA_{0}C_{2}]
+eA_{3}UP[C_{3}(\partial_{2}S_{0})-C_{2}(\partial_{3}S_{0})]\nonumber
\end{eqnarray}
\begin{eqnarray}
&&+UReA_{3}[C_{4}(\partial_{2}S_{0})-C_{2}(\partial_{4}S_{0})]=0,\label{t3}\\
&&-(JP-M^{2})[C_{3}(\partial_{0}S_{0})^{2}-C_{0}(\partial_{0}S_{0})(\partial_{3}S_{0})+
eA_{0}C_{3}(\partial_{0}S_{0})-eA_{3}C_{0}(\partial_{0}S_{0})]\nonumber\\
&&-(JR-MN)[C_{4}(\partial_{0}S_{0})^{2}-C_{0}(\partial_{0}S_{0})(\partial_{4}S_{0})
+eA_{0}C_{4}(\partial_{0}S_{0})]-NU[C_{2}(\partial_{0}S_{0})\nonumber\\
&&(\partial_{4}S_{0})-C_{4}(\partial_{0}S_{0})(\partial_{2}S_{0})
]-(MR-NP)[C_{4}(\partial_{0}S_{0})(\partial_{3}S_{0})-C_{3}(\partial_{0}S_{0})(\partial_{4}S_{0})]\nonumber\\
&&-TM[C_{0}(\partial_{1}S_{0})^{2}-C_{1}(\partial_{0}S_{0})(\partial_{1}S_{0})-
eA_{0}C_{1}(\partial_{1}S_{0})]-TP[C_{3}(\partial_{1}S_{0})^{2}-C_{1}\nonumber\\
&&(\partial_{3}S_{0})(\partial_{1}S_{0})-eA_{3}C_{1}(\partial_{1}S_{0})]
-TR[C_{4}(\partial_{1}S_{0})^{2}-C_{1}(\partial_{1}S_{0})(\partial_{4}S_{0})]-MU\nonumber\\
&&[C_{0}(\partial_{2}S_{0})^{2}-C_{2}(\partial_{0}S_{0})(\partial_{2}S_{0})
-eA_{0}C_{2}(\partial_{2}S_{0})]-UP[C_{3}
(\partial_{2}S_{0})^{2}-C_{2}(\partial_{2}S_{0})\nonumber\\
&&(\partial_{3}S_{0})-eA_{3}C_{2}(\partial_{2}S_{0})]-UR[C_{4}(\partial_{2}S_{0})^{2}-
C_{2}(\partial_{2}S_{0})(\partial_{4}S_{0})]+(MR-PN)\nonumber\\
&&[C_{3}(\partial_{0}S_{0})(\partial_{4}S_{0})-C_{0}(\partial_{3}S_{0})
(\partial_{4}S_{0})+eA_{0}C_{3}(\partial_{4}S_{0})-eA_{3}C_{0}(\partial_{4}S_{0})]+PS\nonumber\\
&&[C_{4}(\partial_{4}S_{0})(\partial_{3}S_{0})-C_{3}(\partial_{4}S_{0})+
eA_{3}C_{4}(\partial_{4}S_{0})]-eA_{0}(JP-M^{2})[C_{3}(\partial_{0}S_{0})\nonumber\\
&&-C_{0}(\partial_{3}S_{0})+eA_{0}C_{3}-eA_{3}C_{0}]-eA_{3}(JR-MN)[C_{4}
(\partial_{0}S_{0})-C_{0}(\partial_{4}S_{0})\nonumber\\
&&+eA_{0}C_{4}]-eA_{0}(MR-NP)[C_{4}(\partial_{3}S_{0})-C_{3}(\partial_{4}S_{0})+eA_{3}C_{4}]=0,\label{t4}\\
&&(MN-JR)[C_{0}(\partial_{0}S_{0})(\partial_{3}S_{0})-C_{3}(\partial_{0}S_{0})^{2}
+eA_{3}C_{0}(\partial_{0}S_{0})-eA_{0}C_{3}(\partial_{0}S_{0})]\nonumber\\
&&+(M^{2}-JS)[C_{0}(\partial_{0}S_{0})(\partial_{4}S_{0})-C_{4}(\partial_{0}S_{0})^{2}
-eA_{0}C_{4}(\partial_{0}S_{0})]+NR[C_{3}\nonumber\\
&&(\partial_{0}S_{0})(\partial_{4}S_{0})-C_{4}(\partial_{3}S_{0})(\partial_{0}S_{0})
-eA_{3}C_{4}(\partial_{0}S_{0})]+TN[C_{0}(\partial_{1}S_{0})^{2}-C_{1}\nonumber\\
&&(\partial_{0}S_{0})(\partial_{1}S_{0})-eA_{0}C_{1}(\partial_{1}S_{0})]+TR[C_{3}
(\partial_{1}S_{0})^{2}-C_{1}(\partial_{1}S_{0})(\partial_{3}S_{0})-eA_{3}\nonumber\\
&&C_{1}(\partial_{1}S_{0})]+TS[C_{4}(\partial_{1}S_{0})^{2}
-C_{1}(\partial_{4}S_{0})(\partial_{1}S_{0})]-MN[C_{0}
(\partial_{2}S_{0})^{2}-C_{2}(\partial_{0}S_{0})\nonumber\\
&&(\partial_{2}S_{0})-eA_{0}C_{2}(\partial_{2}S_{0})]-UP[C_{3}
(\partial_{2}S_{0})^{2}-C_{2}(\partial_{2}S_{0})
(\partial_{3}S_{0})-eA_{3}C_{2}(\partial_{2}S_{0})]\nonumber\\
&&-UR[C_{4}(\partial_{2}S_{0})^{2}-C_{2}(\partial_{2}S_{0})(\partial_{4}S_{0})]-
(MR-PN)[C_{3}(\partial_{0}S_{0})(\partial_{3}S_{0})-C_{0}\nonumber\\
&&(\partial_{3}S_{0})^{2}+eA_{0}C_{3}(\partial_{3}S_{0})-eA_{3}C_{0}(\partial_{3}S_{0})]
-PS[C_{4}(\partial_{3}S_{0})^{2}-C_{3}(\partial_{3}S_{0})(\partial_{4}S_{0})\nonumber\\
&&+eA_{3}C_{4}(\partial_{3}S_{0})]+eA_{0}(MN-JR)[C_{0}(\partial_{3}S_{0})-
C_{3}(\partial_{0}S_{0})+eA_{3}C_{0}-eA_{0}C_{3}]\nonumber\\
&&+eA_{0}(M^{2}-JS)[C_{0}(\partial_{4}S_{0})-C_{4}(\partial_{0}S_{0})-eA_{0}C_{4}]
+eA_{0}NR[C_{3}(\partial_{4}S_{0})-C_{4}\nonumber\\
&&(\partial_{3}S_{0})-eA_{3}C_{4}]-eA_{3}(MR-NP)[C_{3}(\partial_{0}S_{0})-C_{0}
(\partial_{3}S_{0})+eA_{0}C_{3}-eA_{3}\nonumber\\
&&C_{0}]-eA_{3}PS[C_{4}(\partial_{3}S_{0})-C_{3}(\partial_{4}S_{0})+eA_{3}C_{4}]=0.\label{t5}
\end{eqnarray}
Using separation of variables technique, we can define the particles
action for this BH as (also defined in Eq.(\ref{RR1}))
\begin{equation}
S_{0}=-(E-\sum_{i=1}^{2}j_{i}\check{\Omega}_{i})t+W(r,\theta)+L\phi+K(\varphi).
\end{equation}
For the above $S_{0}$ the preceding set of Eqs.(\ref{t1})-(\ref{t5})
can be written in terms of matrix equation, i.e.,
$\Lambda(C_{0},C_{1},C_{2},C_{3},C_{4})^{T}=0$, the elements of the
required matrix provide in the following form
\begin{eqnarray}
\Lambda_{00}&=&-JT(\partial_{1}W)^{2}-JU(\partial_{2}W)-(JP-M^{2})L^{2}
-(JP-M^{2})eA_{3}L-(JR\nonumber\\
&&-MN)L(\partial_{3}K)-(JR-MN)L(\partial_{3}K)-(JR-MN)[L(\partial_{3}K)
+eA_{3}\nonumber\\
&&(\partial_{4}K)]-(JS-M^{2})(\partial_{4}K)-m^{2}J-eA_{3}L-e^{2}A^{2}_{3}
-eA_{3}(JR-MN)\nonumber\\
&&(\partial_{4}K)\nonumber\\
\Lambda_{01}&=&-(E-\sum_{i=1}^{2}j_{i}\check{\Omega}_{i})JT(\partial_{1}W)+JTeA_{0}
(\partial_{1}W)+JTeA_{0}(\partial_{1}W)+MT(\partial_{1}W)L\nonumber\\
&&+eA_{3}(\partial_{1}W)+NT(\partial_{1}W)(\partial_{4}K)\nonumber\\
\Lambda_{02}&=&-(E-\sum_{i=1}^{2}j_{i}\check{\Omega}_{i})JU(\partial_{2}W)+JUeA_{0}
(\partial_{2}W)+MU(\partial_{1}W)L+MUeA_{3}(\partial_{2}W)\nonumber\\
&&+NU(\partial_{2}W)(\partial_{4}K)\nonumber\\
\Lambda_{03}&=&-MT(\partial_{1}W)^{2}-MU(\partial_{2}W)^{2}-
(JP-M^{2})(E-\sum_{i=1}^{2}j_{i}\check{\Omega}_{i})L+(JP-M^{2})eA_{0}\nonumber\\
&&L-(MR-NP)L(\partial_{4}K)-(JR-MN)[(E-\sum_{i=1}^{2}j_{i}\check{\Omega}_{i})
(\partial_{4}K)-eA_{0}(\partial_{4}K)]\nonumber\\
&&+NR(\partial_{4}K)^{2}-m^{2}M-(JP-M^{2})[(E-\sum_{i=1}^{2}j_{i}\check{\Omega}_{i})
eA_{3}-e^{2}A_{0}A_{3}]-eA_{3}(M\nonumber\\
&&R-NP)(\partial_{4}K)\nonumber\\
\Lambda_{04}&=&-NT(\partial_{1}W)-NU(\partial_{2}W)-(JR-MN)
[(E-\sum_{i=1}^{2}j_{i}\check{\Omega}_{i})-eA_{0}L]+(MR\nonumber\\
&&-NP)[L^{2}-eA_{3}L]-(JS-M^{2})[(E-\sum_{i=1}^{2}j_{i}\check{\Omega}_{i})
(\partial_{4}K)-eA_{0}(\partial_{4}K)]-NR\nonumber\\
&&[(\partial_{4}K)+eA_{3}(\partial_{4}K)]-m^{2}N-eA_{3}(JR-MN)
[(E-\sum_{i=1}^{2}j_{i}\check{\Omega}_{i})-eA_{0}]+eA_{3}\nonumber\\
&&(MR-NP)[L+eA_{3}]\nonumber\\
\Lambda_{10}&=&-(E-\sum_{i=1}^{2}j_{i}\check{\Omega}_{i})(\partial_{1}W)JT+TM(\partial_{1}W)L
+TN(\partial_{1}W)(\partial_{4}K)-eA_{0}JT(\partial_{1}W)\nonumber\\
&&+TMeA_{3}(\partial_{1}W)\nonumber
\end{eqnarray}
\begin{eqnarray}
\Lambda_{11}&=&-JT[(E-\sum_{i=1}^{2}j_{i}\check{\Omega}_{i})^{2}
-eA_{0}(E-\sum_{i=1}^{2}j_{i}\check{\Omega}_{i})]
-MTL[(E-\sum_{i=1}^{2}j_{i}\check{\Omega}_{i})+eA_{3}]\nonumber\\
&&+NT(E-\sum_{i=1}^{2}j_{i}\check{\Omega}_{i})(\partial_{4}K)-TU(\partial_{2}W)
+TML[(E-\sum_{i=1}^{2}j_{i}\check{\Omega}_{i})-eA_{0}]\nonumber\\
&&-TPL[L+eA_{3}]-TRL(\partial_{4}K)+TN[(E-\sum_{i=1}^{2}j_{i}
\check{\Omega}_{i})(\partial_{4}K)+eA_{0}(\partial_{4}K)]\nonumber\\
&&-TR[(\partial_{4}K)L+eA_{3}(\partial_{4}K)]-TS(\partial_{4}K)
-m^{2}T-eA_{0}JT[(E-\sum_{i=1}^{2}j_{i}\check{\Omega}_{i})\nonumber\\
&&-eA_{0}]-MTeA_{0}[L+eA_{3}]-eA_{0}NT(\partial_{4}K)+TMeA_{3}
[(E-\sum_{i=1}^{2}j_{i}\check{\Omega}_{i})\nonumber\\
&&-eA_{0}]-TP[L+eA_{3}]-eA_{3}TR(\partial_{4}K)\nonumber\\
\Lambda_{12}&=&TU(\partial_{1}W)(\partial_{2}W)\nonumber\\
\Lambda_{13}&=&-(E-\sum_{i=1}^{2}j_{i}\check{\Omega}_{i})
(\partial_{1}W)MT+TP(\partial_{1}W)L+
TR(\partial_{1}W)(\partial_{4}K)+TMeA_{0}\nonumber\\
&&(\partial_{1}W)+TPeA_{3}(\partial_{1}W)\nonumber\\
\Lambda_{14}&=&-(E-j\check{\Omega})(\partial_{1}W)NT+TR(\partial_{1}W)L
+TS(\partial_{1}W)(\partial_{4}K)+eA_{0}NT(\partial_{1}W)\nonumber\\
&&+eA_{3}TR(\partial_{1}W)\nonumber\\
\Lambda_{20}&=&-UJ(E-\sum_{i=1}^{2}j_{i}\check{\Omega}_{i})
(\partial_{2}W)+MUL(\partial_{2}W)
+MU(\partial_{2}W)(\partial_{4}K)-JUeA_{0}\nonumber\\
&&(\partial_{2}W)+MUeA_{3}(\partial_{2}W)\nonumber\\
\Lambda_{21}&=&TU(\partial_{2}W)(\partial_{1}W)\nonumber\\
\Lambda_{22}&=&-UJ((E-\sum_{i=1}^{2}j_{i}\check{\Omega}_{i})
[(E-\sum_{i=1}^{2}j_{i}\check{\Omega}_{i})-eA_{0}]+
MU(E-\sum_{i=1}^{2}j_{i}\check{\Omega}_{i})[L+eA_{3}]\nonumber\\
&&+NU(E-\sum_{i=1}^{2}j_{i}\check{\Omega}_{i})(\partial_{4}K)-TU(\partial_{1}W)+
MUL[(E-\sum_{i=1}^{2}j_{i}\check{\Omega}_{i})-eA_{0}]\nonumber\\
&&-UPL[L+eA_{3}]-URL(\partial_{4}K)+MU(\partial_{4}K)
[(E-\sum_{i=1}^{2}j_{i}\check{\Omega}_{i})-eA_{0}]-UPL\nonumber\\
&&(\partial_{4}K)-UR(\partial_{4}K)-m^{2}U+JUeA_{0}[eA_{0}-
(E-\sum_{i=1}^{2}j_{i}\check{\Omega}_{i})]-MULeA_{0}\nonumber
\end{eqnarray}
\begin{eqnarray}
&&-NUeA_{0}(\partial_{4}K)+MUeA_{3}[(E-\sum_{i=1}^{2}j_{i}\check{\Omega}_{i})-eA_{0}]-UPLeA_{3}
-UReA_{3}\nonumber\\&&(\partial_{4}K)\nonumber\\
\Lambda_{23}&=&-(E-\sum_{i=1}^{2}j_{i}\check{\Omega}_{i})(\partial_{2}W)MU+UPL(\partial_{2}W)+
UP(\partial_{2}W)(\partial_{4}K)+MUeA_{0}\nonumber\\
&&(\partial_{2}W)+eA_{3}UP(\partial_{2}W)\nonumber\\
\Lambda_{24}&=&-NU(E-\sum_{i=1}^{2}j_{i}\check{\Omega}_{i})(\partial_{2}W)+UR(\partial_{2}W)L
+UR(\partial_{2}W)(\partial_{4}K)NUeA_{0}(\partial_{2}W)\nonumber\\
&&+UReA_{3}(\partial_{2}W)\nonumber\\
\Lambda_{30}&=&-(JP-M^{2})(E-\sum_{i=1}^{2}j_{i}\check{\Omega}_{i})[L+eA_{3}]
-(JR-MN)(E-\sum_{i=1}^{2}j_{i}\check{\Omega}_{i})(\partial_{4}K)\nonumber\\
&&-TM(\partial_{1}W)-MU(\partial_{2}W)^{2}-(MR-PN)(\partial_{4}K)[L+eA_{3}]+eA_{0}(JP-\nonumber\\
&&M^{2})[L+eA_{3}]+eA_{0}(JR-MN)(\partial_{4}K)\nonumber\\
\Lambda_{31}&=&-TM(\partial_{1}W)[(E-\sum_{i=1}^{2}j_{i}\check{\Omega}_{i})-eA_{0}]
+TP(\partial_{1}W)[L+eA_{3}]+TR(\partial_{1}W)\nonumber\\&&(\partial_{4}K)\nonumber\\
\Lambda_{32}&=&-MU(\partial_{2}W)[(E-\sum_{i=1}^{2}j_{i}\check{\Omega}_{i})
-eA_{0}]+UP(\partial_{2}W)[L+eA_{3}]+UR(\partial_{2}W)\nonumber\\&&(\partial_{4}K)\nonumber\\
\Lambda_{33}&=&-(JP-M^{2})(E-\sum_{i=1}^{2}j_{i}\check{\Omega}_{i})
[(E-\sum_{i=1}^{2}j_{i}\check{\Omega}_{i})-eA_{0}]-(MR-NP)(E-\nonumber\\
&&\sum_{i=1}^{2}j_{i}\check{\Omega}_{i})(\partial_{4}K)
-TP(\partial_{1}W)+UP(\partial_{2}W)-(MR-PN)(\partial_{4}K)[(E-\nonumber\\
&&\sum_{i=1}^{2}j_{i}\check{\Omega}_{i})-eA_{0}]-PS(\partial_{4}K)+eA_{0}(JP-M^{2})
[(E-\sum_{i=1}^{2}j_{i}\check{\Omega}_{i})-eA_{0}]\nonumber\\&&+eA_{0}(\partial_{4}K)\nonumber\\
\Lambda_{34}&=&-(JR-MN)(E-\sum_{i=1}^{2}j_{i}\check{\Omega}_{i})[(E-\sum_{i=1}^{2}j_{i}
\check{\Omega}_{i})-eA_{0}]+(MR-NP)(E-\nonumber\\
&&\sum_{i=1}^{2}j_{i}\check{\Omega}_{i})L-TR(\partial_{1}W)-UR(\partial_{2}W)+PS(\partial_{4}K)[L+eA_{3}]
+eA_{0}(JR\nonumber
\end{eqnarray}
\begin{eqnarray}
&&-MN)[(E-\sum_{i=1}^{2}j_{i}\check{\Omega}_{i})-eA_{0}]-eA_{0}(MR-NP)[L+eA_{3}]\nonumber\\
\Lambda_{40}&=&-(MN-JR)(E-\sum_{i=1}^{2}j_{i}\check{\Omega}_{i})[L+eA_{3}]
-(M^{2}-JS)(E-\sum_{i=1}^{2}j_{i}\check{\Omega}_{i})\nonumber\\
&&(\partial_{4}K)+TN(\partial_{1}W)^{2}-MN(\partial_{2}W)^{2}+(MR-PN)L[L+eA_{3}]+eA_{0}\nonumber\\
&&(MN-JR)[L+eA_{3}]+eA_{0}(M^{2}-JS)(\partial_{4}K)+eA_{3}(MR-PN)\nonumber\\&&[L+eA_{3}]\nonumber\\
\Lambda_{41}&=&TN(\partial_{1}W)[(E-\sum_{i=1}^{2}j_{i}\check{\Omega}_{i})+eA_{1}]-TR(\partial_{1}W)
[L+eA_{3}]+TS(E-\nonumber\\&&\sum_{i=1}^{2}j_{i}\check{\Omega}_{i})(\partial_{1}W)\nonumber\\
\Lambda_{42}&=&-MN(\partial_{2}W)[(E-j\check{\Omega})-eA_{0}]+UP(\partial_{2}W)
[L+eA_{3}]+UR(\partial_{2}W)\nonumber\\&&(\partial_{4}K)\nonumber\\
\Lambda_{43}&=&-(MN-JR)(E-\sum_{i=1}^{2}j_{i}\check{\Omega}_{i})
[(E-\sum_{i=1}^{2}j_{i}\check{\Omega}_{i})-eA_{0}]-NR(E-\nonumber\\
&&\sum_{i=1}^{2}j_{i}\check{\Omega}_{i})(\partial_{4}K)+TR
(\partial_{1}W)^{2}-UP(\partial_{2}W)^{2}+(MR-PN)L[(E-\nonumber\\
&&\sum_{i=1}^{2}j_{i}\check{\Omega}_{i})-eA_{0}]+PSL(\partial_{4}K)
+eA_{0}(MN-JR)[(E-\sum_{i=1}^{2}j_{i}\check{\Omega}_{i})\nonumber\\&&-eA_{0}]+eA_{0}NR
(\partial_{4}K)+eA_{3}(MR-PN)[(E-\sum_{i=1}^{2}j_{i}\check{\Omega}_{i})-eA_{0}]\nonumber\\
\Lambda_{44}&=&(JS-M^{2})(E-\sum_{i=1}^{2}j_{i}\check{\Omega}_{i})
[(E-\sum_{i=1}^{2}j_{i}\check{\Omega}_{i})-eA_{0}]+NR(E-\sum_{i=1}^{2}\nonumber\\
&&j_{i}\check{\Omega}_{i})[L+eA_{3}]+TS(\partial_{1}W)-UR(\partial_{2}W)^{2}-PSL[L+eA_{3}]\nonumber\\
&&-e^{2}A^{2}_{0}(M^{2}-JS)-eA_{0}NR[L+eA_{3}]\nonumber
\end{eqnarray}
For the non-trivial solution, the determinant $\Lambda$ is equal to
zero provides the following expression
\begin{eqnarray}
ImW^{\pm}&&=\pm\int\sqrt{\frac{(E-eA_{0}-\Omega _{1}eA_{3}
-\sum_{i=1}^{2}j_{i}\check{\Omega}_{i})^{2}+\tilde{X}}{\frac{TR}{MN-JR}}},\nonumber\\
&&=\pm
\iota\pi\frac{(E-eA_{0}-\Omega_{1}eA_{3}-\sum_{i=1}^{2}j_{i}\check{\Omega}_{i})}{2
\kappa(r_{+})},\nonumber
\end{eqnarray}
where
\begin{eqnarray}
X&=&\frac{NR}{MN-JR}(E-\sum_{i=1}^{2}j_{i}\check{\Omega}_{i})
(\partial_{4}K)+\frac{UP}{MN-JR}(\partial_{2}W)^{2}-\nonumber\\
&&\frac{(MR-PN)L}{MN-JR}[(E-\sum_{i=1}^{2}j_{i}\check{\Omega}_{i})-
eA_{0}]-\frac{PSL}{MN-JR}(\partial_{4}K)-\nonumber\\
&&eA_{0}\frac{NR}{MN-JR}(\partial_{4}K)-eA_{3}\frac{(MR-PN)}{MN-JR}
[(E-\sum_{i=1}^{2}j_{i}\check{\Omega}_{i})\nonumber\\
&&-eA_{0}]-e^{2}A_{3}^{2}+2(E-\sum_{i=1}^{2}j_{i}\check{\Omega}_{i})eA_{3}-2e^{2}A_{3}A_{0}\nonumber
\end{eqnarray}
and the surface gravity is
\begin{equation}
{\kappa(r_{+})}=\frac{1}{2}\sqrt{\tilde{M_{r}}(r,\theta)\tilde{N_{r}}(r,\theta)},\nonumber
\end{equation}
where
$-\tilde{B_{r}}(r,\theta)=\tilde{M_{r}}(r,\theta)~\textmd{and}~
\tilde{C^{-1}_{r}}(r,\theta) =\tilde{N_{r}}(r,\theta)$. The required
tunneling probability is
\begin{equation}
\tilde{\Gamma}=\frac{\tilde{\Gamma}_{emission}}{\tilde{\Gamma}_{absorption}}
=\exp[{-4ImW^{+}}]=\exp\left[{-4\pi \frac{(E-eA_{0}-\Omega
_{1}eA_{3}
-\sum_{i=1}^{2}j_{i}\check{\Omega}_{i})}{\sqrt{\tilde{M_{r}}
(r,\theta)\tilde{N_{r}}(r,\theta)}}}\right].\label{W1}
\end{equation}
The Hawking temperature for BH in 5D by using Boltzmann factor
$\beta=\frac{1}{T_{H}}$, is given by
\begin{equation}
T_{H}=\frac{\sqrt{\tilde{M_{r}}(r,\theta)\tilde{N_{r}}(r,\theta)}
}{4\pi}=\frac{\sqrt{2r^{2}[r^{2}_{0}r^{3}-b^{2}-rb^{2}\sin^{2}\theta
[b^{2}+2a^{2}-2r^{2}-1]}}{4\pi(r^{2}+a^{2}\cos^{2}\theta+b^{2}\sin^{2}\theta)
(r^{4}+b^{2}r^{2}+a^{2}r^{2}-r^{2}_{0}r^{2})}.\label{W2}
\end{equation}
The tunneling probability related to $E$, $A_{0}$, $A_{3}$,
$\check{\Omega}$ angular momentum, surface gravity of a BH. The
Hawking temperature depends on  parameters $r_{0}$, $a$ and $b$.

\section{Outlook}

In this paper, we have used Lagrangian equation to investigate the
tunneling of charged particles from electrically charged black ring
and Myers-Perry BHs. In 5D, black rings have many unusual properties
not shared by Myers-Perry BHs with spherical topology, e.g., their
event horizon topology is not spherical for the cases of neutral,
dipole and charged black rings.

For black rings, the tunneling spectrum of scalar particles has
already been discussed by using the Hamilton-Jacobi method
\cite{W3}. While, the Dirac particles tunneling phenomenon for black
ring has also been discussed \cite{R23}. Recently, the anomalous
derivation of Hawking radiation has attempted to recover the Hawking
temperature of black rings via gauge and gravitational anomalies at
the horizon \cite{W4}. As far as I know, till now, there is no
references to report Hawking radiation of charged vector particles
across single electrically charged black ring. So it is interesting
to see if charged bosons tunneling process is still applicable in
such exotic spacetime. In our analysis, we have found that the
tunneling probability given by Eq.(\ref{W3}) depends on vector
potential components, i.e., $A_{0}$ and $A_{1}$, energy, angular
momentum, particle's charge and surface gravity of black ring.
While, the Hawking temperature (\ref{W4}) depends on parameter
$\alpha$, i.e., charge of a black ring. We have found that the
recovered Hawking temperature is same as already obtained in the
literature for various particles.

For Myers-Perry BH, the tunneling spectrum of massive scalar and
vector particles tunneling have been discussed in different
coordinate systems \cite{R26}. They investigated the Hawking
temperature in the Painlev$\acute{e}$ coordinates and in the
corotating frames and showed that the coordinate system do not
affect the Hawking temperature. Here, we have evaluated charged
vector particles tunneling from Myers-Perry BH by solving Proca
equations by applying the WKB approximation to the Hamilton-Jacobi
method. For this 5D BH, the tunneling probability given by
Eq.(\ref{W1}) is associated to the energy, vector potential, angular
momentum and surface gravity of BH. While, the Hawking temperature
(\ref{W2}) depends on parameters $r_{0}$, $a$ and $b$, which is
similar to the temperature as given in \cite{R26}. It is to be noted
that the effect of electromagnetism appear only on the tunneling
probability of these vector particles which tunnel through the
horizon but not on the Hawking temperature.

In this paper, we have found that the Hawking temperature related to
the emission rate is similar for every type of particles, i.e.,
scalars, fermions, vectors (bosons either charged or uncharged). It
is to be mentioned that the obtained Hawking temperature is
independent of the background BH geometry, coordinate system,
particles species and the temperature must be same either emitted
particles are charged or uncharged. The Hawking temperature
calculated recovered by various methods is same and agrees with the
temperature generally calculated in the literature \cite{R29}.
However, the tunneling probably is different for different cases.


\begin{thebibliography}{99}

\bibitem{R1} R.Emparan, H.S. Reall, Phys. Rev. Lett. {\bf 88}, 101101(2002).

\bibitem{R2} Y.K. Lim, E.Teo, \emph{Geodesics Through the Black Ring},
Report, University of Singapore, 2008.

\bibitem{R3} Y. Chen, E. Teo, JHEP {\bf 06}, 068(2012).

\bibitem{R4} M. Matsumoto, H. Yoshino, H. Kodama, Phys. Rev. {\bf D89},
044016(2014).

\bibitem{R5} M.Cvetic, S.S. Gubser, JHEP {\bf 04}, 024(1999).

\bibitem{R6} A.N. Alive, V.P. Frolov, Phys. Rev. {\bf D69}, 084022(2004).

\bibitem{R7} M. Sharif, W. Javed, Eur. Phys. J. {\bf C72}, 1997(2012).

\bibitem{R8} K. Jan, H. Gohar, Astrophys Space Sci. {\bf 350},
279(2014).

\bibitem{R9} S.I. Kruglov, Int. J. Mod. Phys. {\bf A29},
1450118(2014).

\bibitem{R10} G. Li, X. Zu, Int. J. Applied Maths Phys. {\bf 3},
134(2015).

\bibitem{R11} Z. Feng, Y. Chen, X. Zu, Astrophys. Space Sci. {\bf 48},
359(2015).

\bibitem{R12} M. Saleh, B.B. Thomas, T.C. Kofane, Front. Phys. {\bf 10},
100401(2015).

\bibitem{R13} H. Lin, K. Saifullah, S.T. Yau, Mod. Phys. Lett. {\bf A30},
1550044(2015).

\bibitem{R14} G.R. Chen, Y.C. Huang, Int. J. of Mod. Phys. {\bf
A30}, 1550083(2015).

\bibitem{R15} X.Q. Li, G.R. Chen, Phys. Lett. {\bf B751}, 38(2015).

\bibitem{R16} T.I. Singh, I.A. Meitei, K.Y. Singh, Astrophys. Space Sci. {\bf 361},
103(2015).

\bibitem{R17} H. Gursel, I. Sakalli, Canadian J. Phys. {\bf 94},
147(2016).

\bibitem{R18} X.Q. Li, Phys. Lett. {\bf B763}, 80(2016).

\bibitem{R19} A. \"{O}vg\"{u}n , A. Jusufi, Eur. Phys. J. Plus {\bf 177},
131(2016).

\bibitem{R20} R. Li, J.K. Zhao, Commun. Theor. Phys. {\bf 131},
177(2016).

\bibitem{R21} K. Jusufi, Ali \"{O}vg\"{u}n, Astrophys. Space Sci.{\bf 361},
207(2016).

\bibitem{R23} Q.Q. Jiang, Phy. Rev. {\bf D78}, 044009(2008).

\bibitem{R24} T. Shivalingaswamy, B.A. Kagali, European J. Physics Education  {\bf 2}, 1309(2011).

\bibitem{R25} J.B. Griffith; \emph{Colliding Plane Waves in General Relativity}
(Oxford University Press, 1991).

\bibitem{R26} K. Jusufi, A. \"{O}vg\"{u}n, arXiv:1610.07069v2.

\bibitem{R27} S. Abdolrahimi, J. Kunz, P. Nedkova, Phys. Rev. {\bf D91}, 064068(2015).

\bibitem{W1} R.C. Myers and M.J. Perry, Ann. Phys. {\bf 172}, 304(1986).

\bibitem{R28} L.A. Lopez, Rev. Mex. Fix. {\bf 60}, 95(2014).

\bibitem{W2} F.R. Tangherlini, Nuovo Cim. {\bf 27}, 636(1963).

\bibitem{W3} L. Zhao, arXiv: Commun. Theor. Phys. {\bf 47}, 835(2007).

\bibitem{W4} U. Miyamoto and K. Murata, Phys. Rev. {\bf D77}, 024020(2008); B.
Chen and W. He, Classical Quantum Grav. {\bf 25}, 135011(2008).

\bibitem{R29} A. Yale, Phys. Lett. {\bf B697}, 398(2011).

\end{thebibliography}
\end{document}